\documentclass[conference, 9pt]{IEEEtran}
\IEEEoverridecommandlockouts

\usepackage{cite}
\usepackage{amsmath,amssymb,amsfonts}
\usepackage{graphicx}
\usepackage{textcomp}
\usepackage{ dsfont }
\usepackage{tablefootnote}
\usepackage{pbox}
\usepackage{enumitem}
\usepackage{pifont}
\usepackage{tikz}
\usepackage{inconsolata}
\usepackage{lipsum}
\usepackage{mathrsfs}
\usepackage{longtable}

\def\BibTeX{{\rm B\kern-.05em{\sc i\kern-.025em b}\kern-.08em
    T\kern-.1667em\lower.7ex\hbox{E}\kern-.125emX}}

\renewcommand{\IEEEbibitemsep}{0pt plus 0.5pt}
\makeatletter
\IEEEtriggercmd{\reset@font\normalfont\fontsize{7.8pt}{8pt}\selectfont}
\makeatother
\IEEEtriggeratref{1}

\newcommand\encircle[1]{
    \tikz[baseline=(X.base)] 
    \node (X) [draw, shape=circle, inner sep=-1, fill=black, text=white] {\strut #1};
}

\usepackage{xspace}
\newcommand{\Design}{\textit{$\mathsf{SeDA}$}}
\newcommand{\cmark}{\ding{51}}%

\usepackage[utf8]{inputenc}
\usepackage[T1]{fontenc}
\usepackage{hyperref}
\usepackage{url}
\usepackage{booktabs}
\usepackage{nicefrac}
\usepackage{microtype}
\usepackage{xcolor}
\usepackage{colortbl}
\usepackage{subfigure}
\usepackage{svg}
\usepackage{multirow}
\usepackage{wrapfig}
\usepackage{threeparttable}
\usepackage{makecell}

\usepackage{algorithmic}
\usepackage[linesnumbered,ruled,vlined]{algorithm2e}
\hypersetup{
	colorlinks=true,
	linkcolor=black,
	filecolor=black,      
	urlcolor=teal,
	citecolor=black,
}

\begin{document}
\title{SeDA: \underline{S}ecure and \underline{E}fficient \underline{D}NN \underline{A}ccelerators with Hardware/Software Synergy\vspace{-3mm}}

\author{
    \IEEEauthorblockN{Wei Xuan$^\dag$$^\ddagger$, Zhongrui Wang$^\star$, Lang Feng$^\S$, Ning Lin$^\P$, Zihao Xuan$^\dag$$^\ddagger$, Rongliang Fu$^\parallel$, \\Tsung-Yi Ho$^\parallel$, Yuzhong Jiao$^\dag$$^\ddagger$, Luhong Liang$^\dag$$^\ddagger$}
    \IEEEauthorblockA{\textit{$^\dag$ACCESS – AI Chip Center for Emerging Smart Systems, InnoHK Centers, Hong Kong Science Park, Hong Kong, China}\\
    \textit{$^\ddagger$The Hong Kong University of Science and Technology, Hong Kong, China}\\
    \textit{$^\star$Southern University of Science and Technology, Shenzhen, Guangdong, China}\\
    \textit{$^\S$Sun Yat-sen University Shenzhen Campus, Shenzhen, Guangdong, China}\\
    \textit{$^\P$The University of Hong Kong, Hong Kong, China}\\
    \textit{$^\parallel$The Chinese University of Hong Kong, Hong Kong, China}
    \\Corresponding authors: wangzr@sustech.edu.cn, flang1994@outlook.com}
}
\maketitle

\begin{abstract}

Ensuring the confidentiality and integrity of DNN accelerators is paramount across various scenarios spanning autonomous driving, healthcare, and finance. However, current security approaches typically require extensive hardware resources, and incur significant off-chip memory access overheads. This paper introduces $\Design$, which utilizes 1) a \textit{bandwidth-aware encryption mechanism} to improve hardware resource efficiency, 2) \textit{optimal block granularity} through intra-layer and inter-layer tiling patterns, and 3) a \textit{multi-level integrity verification mechanism} that minimizes, or even eliminates, memory access overheads. Experimental results show that $\Design$ decreases performance overhead by over 12\% for both server and edge neural processing units (NPUs), while ensuring robust scalability. \footnote{$\Design$ source code: \url{https://github.com/wayne4s/seda.git}}

\end{abstract}

\section{Introduction}\label{sec:intro}

Securing Deep Neural Networks (DNNs) on neural processing units (NPUs) is increasingly vital for mission-critical applications in areas such as autonomous driving~\cite{parekh2022review}, healthcare~\cite{rong2020artificial}, and finance~\cite{rundo2019machine}. The Artificial Intelligence (AI) hardware market was valued at USD 54.10 billion in 2023 and is expected to surge to USD 474.10 billion by 2030, reflecting a remarkable CAGR of $38.73\%$~\cite{aihwmarket}. Protecting DNN accelerators by ensuring confidentiality and integrity  is crucial for several reasons: \textit{Data Confidentiality}: The sanctity of training data is non-negotiable. Protecting training data is essential to prevent unauthorized access and exploitation of private or sensitive information. \textit{Resource Costliness}: The substantial investment in training resources necessitates stringent security protocols. Safeguarding these resources not only ensures their optimal utilization but also protects against financial losses and inefficiencies. \textit{Vulnerability to Malicious Attacks}: Malicious intent poses a significant threat to DNN models. Protecting these models from potential tampering and attacks is critical to maintaining their integrity and functionality, safeguarding against adverse outcomes and ensuring the reliability of AI systems.

\begin{figure}
    \centering
    \includegraphics[width=0.99\linewidth]{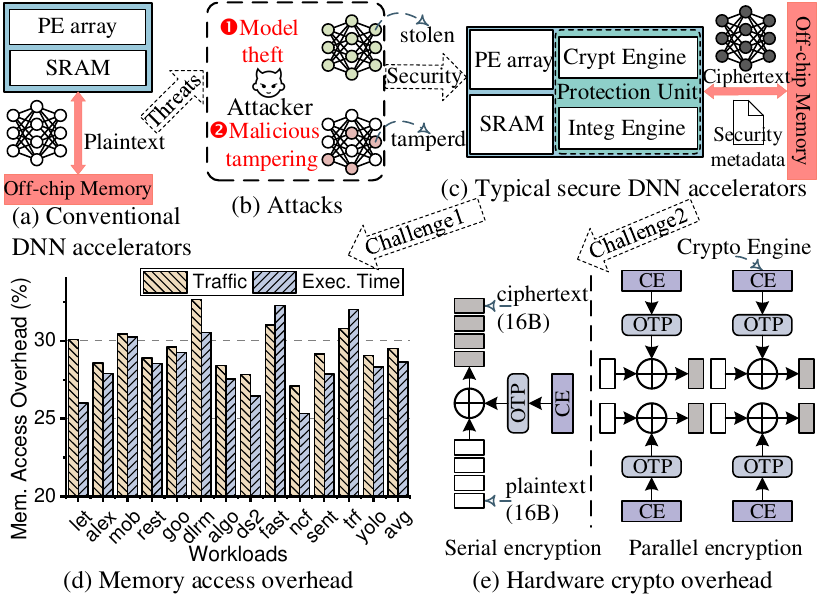}
    \caption{Insight of typical secure accelerators. (a) Traditional DNN accelerators use untrusted off-chip memory and communication buses, risking model theft and malicious tampering, as shown in (b). (c) Secure DNN accelerators typically protect data confidentiality and integrity with memory protection schemes using AES encryption and Hash function. (d) Accessing security metadata in off-chip memory adds overhead. (e) Parallel hardware encryption incurs overhead to meet bandwidth needs of accelerators.}
    \label{figs:intro_overview}
\end{figure}

To protect traditional DNN accelerators from model theft and malicious tampering, as illustrated in Fig.~\ref{figs:intro_overview}(a) and Fig.~\ref{figs:intro_overview}(b), broad approaches~\cite{costan2016intel,zuo2021sealing, hua2022guardnn, hua2022mgx, lee2022tnpu, lee2023secureloop, shrivastava2023securator} have been proposed to secure these accelerators, with the primary aim of minimizing off-chip memory access overhead for security metadata. They typically use the counter-mode encryption of Advanced Encryption Standard (AES-CTR) for confidentiality and the message authentication code (MAC) for integrity verification, as shown in Fig.~\ref{figs:intro_overview}(c). The counter value in AES-CTR concatenates the physical address (PA) and a \textcircled{a}~version number (VN) of a data block, with the VN stored off-chip and incremented with each write. To ensure the integrity of off-chip memory data, each data block is accompanied by a \textcircled{b}~message authentication code (MAC). Additionally, a \textcircled{c}~Merkle Tree (MT)~\cite{gassend2003caches} and its variants are often utilized, with the root stored on-chip to prevent replay attacks.

\textbf{Motivation}~\encircle{1} $\blacktriangleright$ \textit{Costly Off-Chip Memory Access Overhead for Integrity Verification}. Accessing security metadata (e.g., \textcircled{a}\textcircled{b}\textcircled{c})  to ensure the confidentiality and integrity of untrusted off-chip memory significantly increases memory access overhead, as shown in Fig.~\ref{figs:intro_overview}(d). Existing approaches propose several optimizations: using Bonsai Merkle Tree (BMT)~\cite{rogers2007using} instead of traditional MT for smaller version numbers (VNs); employing coarser-grained protection units, such as 512B data blocks instead of 64B cachelines; and dynamically updating VNs based on DNN model state information to eliminate off-chip memory access~\cite{hua2022mgx,lee2022tnpu}. Securator~\cite{shrivastava2023securator} uses a layer-level MAC to reduce off-chip memory access for integrity checks. However, it overlooks the overlap of intra-layer tiles and distinct tiling patterns across layers, potentially leading to redundant encryption/decryption and integrity verification overhead. Moreover, not considering inter-layer tiling pattern differences may result in false negatives.

\textbf{Motivation}~\encircle{2} $\blacktriangleright$ \textit{High Hardware Overhead for Confidentiality Protection}. Due to the AES Engine's limitation of en/decrypting only a 128-bit data block at one time, typical solutions to meet the high bandwidth demands of DNN accelerators involve increasing the number of AES Engines. For example, Securator~\cite{shrivastava2023securator} uses four AES engines to en/decrypt a 64B data block, as shown in Fig.~\ref{figs:related_work_aesctr}(c). However, this approach adds strain on resource-limited accelerators, requiring a careful balance between hardware resource allocation and security performance. As illustrated in Fig.~\ref{figs:intro_overview}(e), a single Crypto Engine can only perform serial encryption, failing to meet the high bandwidth demands of DNN accelerators, while multiple AES engines can satisfy these demands but result in significant hardware encryption overhead.

In this paper, we introduce $\Design$, a novel secure and efficient DNN accelerator architecture with reduced hardware resource consumption and near-zero performance overhead for integrity verification, while maintaining the same level of security and practical applicability. This paper makes the following major contributions:

\begin{itemize}[leftmargin=*]
    \item {\textbf{Insight.} Providing an in-depth insight of limitations of current memory protection strategies for DNN accelerators highlights two critical concerns: the substantial \textit{hardware overhead for encryption} and the expensive \textit{off-chip memory access for integrity verification}.
    }
    \item {\textbf{Solution.} Through hardware/software synergy, we present $\Design$, a secure and efficient accelerator architecture. It incorporates a \textit{bandwidth-aware encryption mechanism} that utilizes a single AES engine with adjustable encryption granularity, minimizing hardware resource overhead. Furthermore, its \textit{multi-level integrity verification mechanism} significantly reduces or eliminates off-chip memory access overhead.
    }
    \item {\textbf{Evaluation.} Conducting extensive experiments using cycle-accurate simulators for DNN accelerators, memory protection schemes, and off-chip memory accesses.  Experimental results of $\Design$ demonstrate a performance improvement, reducing overhead by $12.26\%$ for server NPU and $12.29\%$ for edge NPU, while also providing robust scalability with minimal hardware overhead to meet the bandwidth demands of accelerators.
    }
\end{itemize}

This paper is structured as follows: Section~\ref{sec:related_work} reviews related work on memory protection and the threat model for our study. In Section~\ref{sec:architecture}, we provide a detailed introduction to the proposed $\Design$ architecture. We then conduct extensive experiments on various DNN models in Section~\ref{sec:evaluation}. Finally, Section~\ref{sec:conclusion} concludes our work.
\section{Related Work and Threat Model} \label{sec:related_work}

Based on the corresponding threat model, secure DNN accelerators typically provides robust confidentiality and integrity guarantees in untrusted environments.

\subsection{Confidentiality Protection}

\begin{figure}
    \centering
    \includegraphics[width=0.94\linewidth]{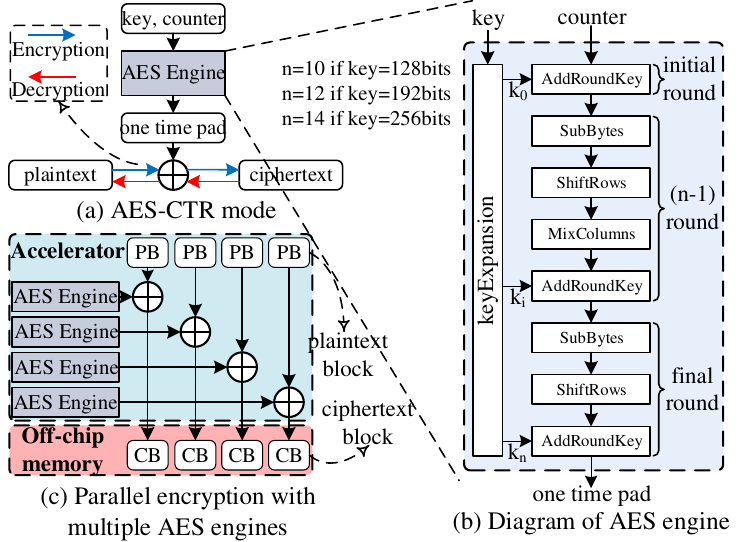}
    \caption{Summary of AES-CTR mode. (a) Reusing AES engine for encryption and decryption in AES-CTR mode. (b) Diagram of the AES engine, featuring the AddRoundKey, SubBytes, ShiftRows, MixColumns, and keyExpansion modules. (c) Utilization of multiple AES engines for parallel encryption to boost high-bandwidth capabilities.}
    \label{figs:related_work_aesctr}
\end{figure}

Confidentiality protection typically uses AES-CTR mode due to several advantages: 1) it utilizes the same AES engine for en/decryption, reducing hardware overhead (see Fig.~\ref{figs:related_work_aesctr}(a)); 2) One time pad (OTP)  generation can be parallelized with communication and DNN computation, minimizing time overhead; and 3) its block-based streaming mode does not require prior knowledge of data size and scale. The AES-CTR encryption requires a non-repeating and incrementing counter to produce a OTP for each encryption/decryption under the same key. The counter concatenates the physical address of a data block that will be encrypted and version number that is incremented on each write memory operation. Here, let $K_e, \mathcal{P}, \mathcal{C}$ be the AES encryption key, plaintext, ciphertext, respectively. The AES-CTR mode for encryption and decryption can be formulated as following Eq.~\ref{equ:aes_ctr_enc} and Eq.~\ref{equ:aes_ctr_dec}, where $AES\mbox{-}CTR_{K_e}(PA || VN)$ produces the OTP, $||$ and $\oplus$ represents a bit-wise concatenation and XOR operator, respectively.

\vspace{-0.5cm}
\begin{align}
    \mathcal{C} &= \mathcal{P} \oplus AES\mbox{-}CTR_{K_e}(PA \; || \; VN) \label{equ:aes_ctr_enc} \\
    \mathcal{P} &= \mathcal{C} \, \oplus AES\mbox{-}CTR_{K_e}(PA \; || \; VN) \label{equ:aes_ctr_dec}
\end{align}
\vspace{-0.5cm}

When an accelerator reads a data block from off-chip memory, the protection module uses the VN to generate an OTP and decrypts the block by XORing it with the OTP (red path in Fig.~\ref{figs:related_work_aesctr}(a), Eq.~\ref{equ:aes_ctr_dec}). Similarly, when writing a data block to off-chip memory, the module increments the VN to generate an OTP and encrypts the block by XORing it with the OTP (blue path in Fig.~\ref{figs:related_work_aesctr}(a), Eq.~\ref{equ:aes_ctr_enc}). Fig.~\ref{figs:related_work_aesctr}(b) illustrates the diagram of the AES engine along with its key functional modules.

\subsection{Integrity Verification}
To ensure data integrity against off-chip memory tampering, an integrity verification engine computes a message authentication code (MAC) by concatenating a data block with its physical address and version number. This MAC is generated during every write operation and verified during reads to ensure data authenticity. Relying solely on MAC for a data block fails to guarantee freshness and opens the door to replay attacks. To counter this threat, prior works utilize an Integrity Tree for hierarchical MAC verification, such as MT~\cite{gassend2003caches} and Bonsai Merkle Tree (BMT)~\cite{rogers2007using}, with the root stored on-chip to prevent malicious tampering. The overhead of integrity verification is non-trivial since it requires traversing both MACs and the nodes of the Integrity Tree stored in the off-chip memory to prevent the replay attack.

\begin{figure*}
    \centering
    \includegraphics[width=0.98\linewidth]{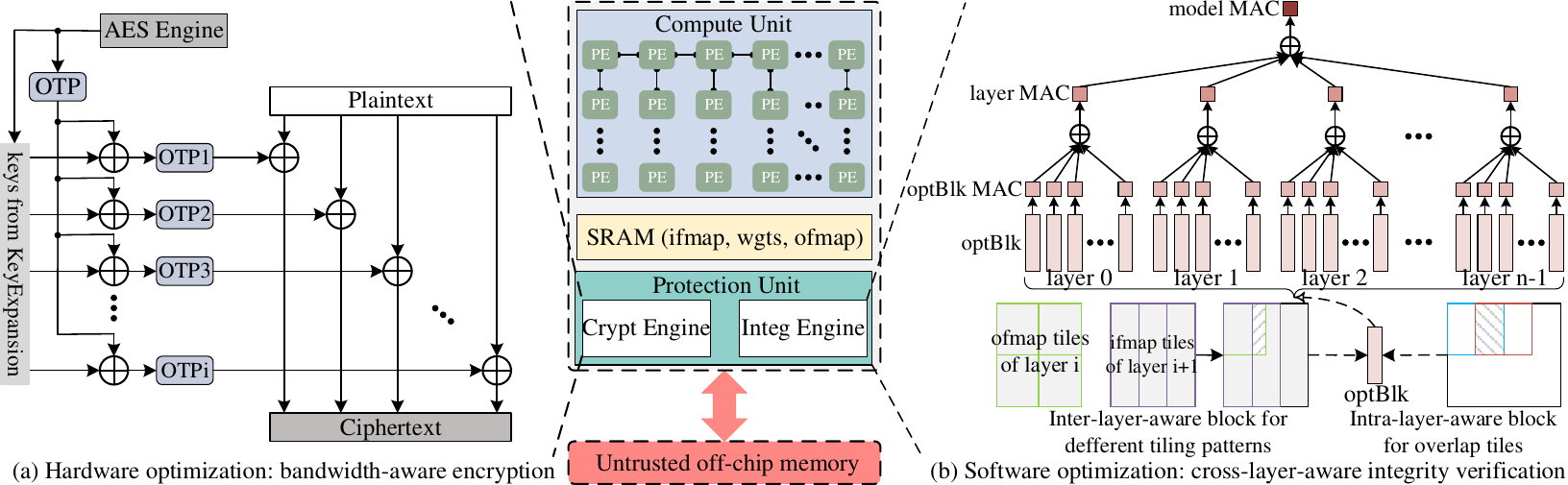}
    \caption{Overview of $\Design$ architecture. (a) The Crypt Engine optimizes hardware by XORing keys from the AES Engine's KeyExpansion module with OPT, creating bandwidth-sensitive encryption granularity and reducing hardware overhead compared to using multiple AES Engines. (b) The Integ Engine optimizes software by analyzing overlapping tiles within a layer and patterns across layers to determine the optimal block size, optBlk, for integrity verification. This leads to a multi-level integrity verification mechanism with optBlk MAC, layer MAC, and model MAC.}
    \label{figs:archi_seda}
\end{figure*}

\subsection{Secure DNN Accelerators}
Intel SGX~\cite{costan2016intel} creates a secure enclave using CPU hardware mechanisms, using AES-CTR mode and MT with its root in the Trusted Computing Base (TCB). SEAL~\cite{zuo2021sealing} introduces a criticality-aware smart encryption scheme that selectively bypasses the encryption engine for partial data and colocates data with corresponding counters. GuardNN~\cite{hua2022guardnn} proposes a secure accelerator architecture by using small TCB and low overhead for privacy-preserving deep learning. MGX~\cite{hua2022mgx} introduces a secure DNN accelerator with application-specific version number management, utilizing on-chip status for version number generation and coarse-grained integrity verification to reduce off-chip memory overhead. In a similar vein, TNPU~\cite{lee2022tnpu} generates all version numbers within an on-chip cache. Securator~\cite{shrivastava2023securator} introduces a layer-level integrity checks by XORing all message authentication codes (MACs) within a layer. 

\subsection{Threat Model}
Similar to the threat model proposed in literature~\cite{hua2022mgx, lee2022tnpu}, we assume that the accelerator itself is secure, meaning that its internal state cannot be directly observed or tampered with. Any external device connected to the accelerator is considered untrusted, such as off-chip memory and communication buses, meaning that attackers could potentially access or manipulate any information stored in these devices through physical or software attacks. Notably, we do not consider other physical side-channel attacks to infer DNN model architectures~\cite{yan2020cache}, such as power channel attacks and electromagnetic attacks. Additionally, adversarial attacks~\cite{akhtar2018threat} against machine learning algorithms are also not within scope to ensure that decryption is restricted to authorized users possessing the key.
\section{$\Design$ Architecture} \label{sec:architecture}
This section initially provides an overview of $\Design$. We then subsequently explores the scalability challenges of cryptographic hardware and the expensive off-chip memory access overhead in integrity verification, providing corresponding solutions for each.

\subsection{Overview}

$\Design$ provides a secure and efficient environment for DNN accelerators by hardware/software co-optimization. The overview architecture of the framework is presented in Fig.~\ref{figs:archi_seda}. The AES Engine's limitation to encrypting only a 128-bit data block at one time poses a challenge in meeting the high bandwidth demands of DNN accelerators. $\Design$ addresses these issues by utilizing bandwidth-aware encryption mechanism to optimize hardware resources impacted by security concerns, as illustrated in Fig.~\ref{figs:archi_seda}(a). Additionally, the inclusion of security metadata by secure DNN accelerators results in storage and access overheads in off-chip memory. For example, using an 8B MAC to represent a 64B data block alone leads to a $12.5\%$ increase in memory traffic. To mitigate this, $\Design$ introduces a multi-level integrity verification software optimization mechanism, showcased in Fig.~\ref{figs:archi_seda}(b). Furthermore, securely housing these limited MACs directly in on-chip SRAM can eliminate off-chip memory access overheads entirely.

\begin{algorithm}[t]
    \DontPrintSemicolon
    \SetKwInOut{Input}{\textbf{Input}}	
    \SetKwInOut{Output}{\textbf{Output}}	
    \SetKwFunction{Attack}{\textbf{Attack of SECA}}{}{}
    \SetKwFunction{Defense}{\textbf{Defense of SECA}}{}{}
    \let\oldnl\nl
    \newcommand{\nonl}{\renewcommand{\nl}{\let\nl\oldnl}}
    \nonl\Attack  \\
    
    \Input{OTP: one time pad for a data block ($blk$); $most\_value\_p$: most used plaintext in $blk$.
        }
    \Output{$value\_p$: all plaintext of $blk$ \\
    }
    
    $most\_value\_c \leftarrowtail $ \textsc{CalcFreqValue}($blk$) \label{alg:seca:attack:calc} \\
    OTP$ \leftarrowtail most\_value\_p \oplus most\_value\_c$ \label{alg:seca:attack:otp} \\
    \For{each encrypted element value\_c of blk} { \label{alg:seca:attack:for:begin}
        $value\_p \leftarrowtail value\_c \; \oplus$ OTP \label{alg:seca:attack:for:end} \\
    }
      
    \vspace{2mm}
    \nonl\Defense \\
    \Input{PA, VN: physical address and version number of $blk$. 
        }
    \Output{OTP$_i$: generate multiple OTPs for $blk$\\
    }
    OTP $\leftarrowtail$ AES-CTR$_{K_e}$(PA || VN) \label{alg:seca:defense:otp} \\
    \For{each 128$\mbox{-}$bit $key_i$ in keyExpansion of AES-CTR} { \label{alg:seca:defense:for:begin}
            OTP$_i \leftarrowtail$  (OTP $\oplus \; key_i) $ \label{alg:seca:defense:for:end}	 \\
    }
    
    \caption{Attack and defense of SECA} \label{alg:seca}
\end{algorithm}

\subsection{Bandwidth-aware encryption mechanism}

To meet the high bandwidth demands of accelerators, increasing the number of AES engines incurs significant overhead, while using the same OTP for all data in a block introduces security risks.This dilemma often leads to a suboptimal trade-off between security and computational efficiency.  

\textbf{Challenge 1: High Hardware Resources for Confidentiality Protection.} While maintaining an equivalent level of security, numerous approaches have bolstered encryption parallelism by stacking multiple AES engines. For example, Securator~\cite{shrivastava2023securator} uses four AES-128 engines to concurrently encrypt a 64B data block, as shown in Fig.~\ref{figs:related_work_aesctr}(c). Nevertheless, the drawback of sacrificing hardware resources to achieve performance improvements in encryption and decryption is apparent, particularly for edge DNN accelerators constrained by limited hardware capabilities. This situation  emphasizes the crucial equilibrium needed to optimize security and operational efficiency in these systems.

\textbf{Challenge 2: Security Threat for Shared OTP.}
A straightforward approach is to use each engine once per data block, with each 128-bit segment within this data block sharing the same OTP. Nevertheless, assigning a OTP to a individual data block poses security risks and could be vulnerable to a \underline{S}ingle-\underline{E}lement \underline{C}ollision \underline{A}ttack (\textbf{SECA}). The principle of a SECA attack when a data block shares the same OTP is outlined in lines~\ref{alg:seca:attack:calc}-\ref{alg:seca:attack:for:end} of Algorithm~\ref{alg:seca}. By employing the \textsc{CalcFreqValue} function, we can identify the most frequently encrypted data within the block, denoted as $most\_value\_c$. Through comprehensive data analysis, we can infer the most common plaintext data within the block, denoted as $most\_value\_p$ (e.g., $0$). Referring to the AES-CTR encryption formula Eq.~\ref{equ:aes_ctr_enc}, we can calculate the OTP for that block as $most\_value\_c \oplus most\_value\_p$, as shown in line~\ref{alg:seca:attack:otp} of Algorithm~\ref{alg:seca}. Since the block shares this OTP, once obtained, we can derive all plaintext values within the encrypted block, as shown in lines~\ref{alg:seca:attack:for:begin}-\ref{alg:seca:attack:for:end} of Algorithm~\ref{alg:seca}. Extending this principle further, we can extract the data values from every DNN layer and even the entire model.

\textbf{Solution.}
We introduce a novel bandwidth-aware encryption mechanism to meet the high bandwidth requirements of accelerators without compromising the security of encryption. This method leverages the inherent features of AES-CTR encryption mode, using just a single AES engine and a minimal number of XOR logic gates, as illustrated in Fig.~\ref{figs:archi_seda}(a). To defend against SECA attacks, we first employ a standard AES-CTR encryption engine to create a shared OTP for a data block, as shown in line~\ref{alg:seca:defense:otp} of Algorithm~\ref{alg:seca}. Subsequently, utilizing the keys array produced by the keyExpansion module within the AES-CTR encryption engine, we can generate multiple distinct OTPs by XORing the OTP with keys, as shown in lines~\ref{alg:seca:defense:for:begin}-\ref{alg:seca:defense:for:end} of Algorithm~\ref{alg:seca}. This ensures that each 128-bit data segment within the data block corresponds to a unique OTP, thereby thwarting SECA attacks to safeguard security. Moreover, this mechanism significantly conserves hardware resources by utilizing a small number of XOR logic gates instead of an equivalent number of AES engines, commonly used by traditional methods.

When it comes to securing the entire DNN model, using a single key in the AES-CTR keyExpansion process is typically sufficient to ensure security. We proceed with the assumption that multiple keys are generated through the keyExpansion module using just one key. The keys produced by key expansion are inherently secure. These keys can be stored on-chip or generated in real-time. By XORing the shared OTP from a data block with these different keys, new secure OTPs can be generated. When generating OTPs from multiple keys derived from a single key input for keyExpansion doesn't meet the accelerator's bandwidth requirements, expanding the key expansion input to $key \; \oplus$ (PA || VN) can solve the issue. This approach produces enough OTPs for a data block, meeting both security and bandwidth needs simultaneously.

\begin{algorithm}[t]
    \DontPrintSemicolon
    \SetKwInOut{Input}{\textbf{Input}}	
    \SetKwInOut{Output}{\textbf{Output}}
    \SetKwFunction{Attack}{\textbf{Attack of RePA}}{}{}
    \SetKwFunction{Defense}{\textbf{Defense of RePA}}{}{}
    \let\oldnl\nl
    \newcommand{\nonl}{\renewcommand{\nl}{\let\nl\oldnl}}
    \nonl\Attack  \\
    
    \Input{ MAC$_i$: the MAC of a data block in one layer.
        }
    \Output{$plaintext\_e$: error plaintext.\\
    }
    
    SUM\_MAC $\leftarrowtail \sum_{i = 1}^{n} \oplus$ MAC$_i$ \label{alg:repa:attack:sum_mac}\\
    $\textsc{ShuffleOrder}($MACs$)$ \label{alg:repa:attack:shuffle} \\
    SUM\_MAC\_shuffle $\leftarrowtail \sum_{i = 1}^{n} \oplus$ MAC$_i^{'}$ \label{alg:repa:attack:shuffle_sum}\\

    \If {True = $\textsc{VerifyInteg}($SUM\_MAC, SUM\_MAC\_suffle$)$} { \label{alg:repa:attack:verify}
        \For{each encrypted data block $blk$ of one layer} { \label{alg:repa:attack:for:begin}			
            $plaintext\_e \leftarrowtail \textsc{Decrypt}(blk)$ \label{alg:repa:attack:for:end} \\
        }
    }
    
    \vspace{2mm}
    \nonl\Defense \\
    \Input{$layer_{id}$, $fmap_{idx}$, $blk_{idx}$: layer number, feature map index and block index of $layer_{id}$. \\ 
    }
    \Output{Secure layer MAC.\\}
    
    \For{each encrypted data block $blk$ of $layer_{id}$} { \label{alg:repa:defense:for:begin}		
        MAC$_i \leftarrowtail \textsc{Hash}_{K_h}(blk$ || PA || VN || $layer_{id}$ || $fmap_{idx}$ || $blk_{idx}$) \label{alg:repa:defense:for:end} \\
    }
    \caption{Attack and defense of RePA} 
    \label{alg:repa}
\end{algorithm}

\subsection{Multi-level integrity verification mechanism}
The granularity of integrity verification, whether too large or too small, can be problematic. Small granularity increases security metadata and off-chip memory overhead, affecting performance. Larger granularity reduces metadata overhead but can lead to redundant processing due to tiling overlaps. $\Design$ addresses this with a multi-level integrity verification mechanism that combines the flexibility of fine granularity, which avoids redundant security computations, with near-zero overhead of coarse granularity, significantly reducing or eliminating off-chip memory access overhead.

\textbf{Challenge 1: Expensive Off-Chip Memory Access for Integrity Checks.}
While recent research efforts~\cite{hua2022mgx, lee2022tnpu} have successfully eliminated the off-chip memory access overhead of VNs and Merkle Trees, the overhead introduced by MACs still remains to be adequately addressed. Securator~\cite{shrivastava2023securator} proposed a layer-level freshness and integrity check method to reduce the MACs overhead by XORing MACs of all blocks within a layer, with a block size granularity of 32 bytes. The approach didn't consider tile overlaps within a layer, causing redundant computations like repeated integrity checks, which adds costs. Securator also ignored different tiling patterns between layers. Fig.~\ref{figs:archi_seda}(b) shows that $ofmap$ in layer i and $ifmap$ in layer i+1 use different strategies, which can vary in size and direction. Not addressing inter-layer tiling properly can lead to extra costs or disrupt the model due to inaccurate integrity checks.

\textbf{Challenge 2: Security Threat for XOR-MAC Scheme.}
The XOR-MAC scheme offers parallelizability, incrementality, and provable security, with its security being on par with that of chaining MAC~\cite{bellare1995xor}. However, XORing all MACs generated by directly hashing the ciphertext within a layer to produce a unique layer MAC could lead to a \underline{Re}-\underline{P}ermutation \underline{A}ttack (RePA), as illustrated in lines~\ref{alg:repa:attack:sum_mac}-\ref{alg:repa:attack:for:end} of Algorithm~\ref{alg:repa}. \textsc{ShuffleOrder} rearranges the order of data blocks in the current layer and computes SUM\_MAC\_shuffle for that layer using XOR operations. Because XOR is commutative, meaning the order of operands doesn't affect the result, an attacker can successfully pass the \textsc{VerifyInteg}(SUM\_MAC, SUM\_MAC\_shuffle) verification. This rearrangement of data blocks within a layer can hinder correct decryption of ciphertext, posing a security risk due to RePA vulnerability.

\begin{table}[t]
    \caption{Comparison of Multi-level Integrity Verification Granularity.}
    \label{table:macs}
    \centering
    \footnotesize
    \renewcommand{\arraystretch}{0.85}
    \setlength{\tabcolsep}{1.8pt}
    \begin{tabular}{c|c|c|c}
        \toprule
        \multicolumn{1}{c|}{\textbf{Granularity}}   & \multicolumn{1}{c|}{\textbf{Flexibility}} & \multicolumn{1}{c|}{\textbf{Off-chip Access Overhead}} & \multicolumn{1}{c}{\textbf{Storage}} \\ 
        \midrule
        \textbf{optBlk}   & \begin{tikzpicture} \fill[gray!100] (0,0) circle (0.15cm); \end{tikzpicture}   &  \begin{tikzpicture} \draw (0,0) circle (0.15cm); \fill[orange!50] (0,0) -- (0.15,0) arc (0:60:0.15cm) -- cycle; \end{tikzpicture} & Off-chip\\
        \rowcolor[HTML]{EFEFEF}
        \textbf{layer}        & \begin{tikzpicture} \fill[gray!60] (0,0) circle (0.15cm); \end{tikzpicture}   & \begin{tikzpicture}  \draw (0,0) circle (0.15cm); \fill[orange!50] (0,0) -- (0.15,0) arc (0:20:0.15cm) -- cycle; \end{tikzpicture} & Off/On-chip \\
        \textbf{model}        &  \begin{tikzpicture} \fill[gray!30] (0,0) circle (0.15cm); \end{tikzpicture} &   \begin{tikzpicture} \draw (0,0) circle (0.15cm); \fill[orange!50] (0,0) -- (0.15,0) arc (0:0:0.15cm) -- cycle; \end{tikzpicture}   &  On-chip \\
        \bottomrule
    \end{tabular}
    \begin{flushleft}
        \small
         *Note: Color depth shows intensity, while coverage area indicates ratio.
    \end{flushleft}
\end{table}

\textbf{Solution.} To reduce off-chip memory access overhead from integrity verification, we propose a secure multi-level mechanism. It combines block-level and layer-level checks to minimize memory access by leveraging deterministic memory access patterns of DNNs, while ensuring security. We use the scheduling search strategy proposed in the SecureLoop~\cite{lee2023secureloop} to obtain the optimal authentication block (optBlk), which synergizes with our work orthogonally. 
We propose a novel multi-level integrity verification mechanism involving three granularities of MAC types: optBlk MAC, layer MAC, and model MAC, as shown in Fig.~\ref{figs:archi_seda}(b). Table~\ref{table:macs} compares their features in detail. The optBlk MAC offers flexibility by accounting for intra-layer tiling overlap and diverse inter-layer tiling patterns, avoiding redundant computations from repeated integrity checks. By XORing all optBlk MACs within a layer to create the layer MAC, we achieve a significant reduction in the number of MACs needed for DNN model integrity verification, despite a slight delay due to layer-specific checks. This allows layer MACs to be stored in on-chip SRAM, eliminating off-chip memory access costs. The Model MAC uses a single MAC to represent the entire model weights on-chip, further reducing off-chip memory costs and conserving on-chip SRAM, with verification results available only at the end of model inference. To prevent RePA attacks, we associate each encrypted optBlk data block with specific location details like PA, VN, $layer_{id}$, $fmap_{idx}$, and $blk_{idx}$. We then compute the corresponding MAC using hashing, as detailed in lines~\ref{alg:repa:defense:for:begin}-\ref{alg:repa:defense:for:end} of Algorithm~\ref{alg:repa}. In summary, our multi-level integrity verification mechanism can remove off-chip memory overhead from security metadata, ensuring security while maintaining integrity guarantee.
\section{Evaluation} \label{sec:evaluation}

\subsection{Experimental Setup} \label{sec:exp_setup}

\noindent\textbf{Accelerators}. In order evaluate DNN inference behaviours, we use an open-source cycle-level DNN simulator SCALE-Sim2 \cite{samajdar2018scale,samajdar2020systematic} developed by ARM Research to 1) study the inference execution for various DNN models; 2) analyse the performance overhead of different memory protection schemes. The DNN accelerator can generate detailed computation information of systolic array, and DRAM access traces. After obtaining the DRAM traces, we use various memory protection mechanisms to calculate execution time and bandwidth usage, producing the total DRAM traces after running the security simulator. Finally, we use the DRAM simulator Ramulator2~\cite{luo2023ramulator} to simulate the total DRAM access traces.

\noindent\textbf{Configurations}. Table~\ref{table:config} presents the DNN simulation configurations, featuring a server NPU (Google TPU v1) and an edge NPU (Samsung Exynos 990). To balance DNN computation and memory bandwidth, we simulate four 64-bit DDR channels for both the server and edge NPUs.

\begin{table}[t]
    \caption{DNN Simulation Configurations~\cite{jouppi2017datacenter,song20197}.}
    \label{table:config}
    \centering
    \footnotesize
    \renewcommand{\arraystretch}{0.85}
    \setlength{\tabcolsep}{1.8pt}
    \begin{tabular}{c|c|c}
        \toprule
        \textbf{Metrics} & \multicolumn{1}{c|}{\textbf{Server (Google TPU v1)}} & \multicolumn{1}{c}{\textbf{Edge (Samsung Exynos 990)}} \\ 
        \midrule        
        \textbf{PE}              & 256 x 256  in systolic array   & 32 x 32 in systolic array \\
        \rowcolor[HTML]{EFEFEF}
        \textbf{Bandwidth}        & 20 GB/s with 4 channels   & 10 GB/s with 4 channels \\
        \textbf{Frequency}        &  1 GHz &   2.75 GHz   \\
        \rowcolor[HTML]{EFEFEF}
        \textbf{SRAM}             & 24 MB & 480 KB \\
        \textbf{Precision}        & 1-B for per element  & 1-B for per element \\
        \bottomrule
    \end{tabular}
\end{table}

\noindent\textbf{Benchmarks}. For DNN accelerators, we evaluate $\Design$ across various DNN models, including Lenet (let), Alexnet (alex), Mobilenet (mob), ResNet18 (rest), GoogleNet (goo), DLRM (dlrm), AlphaGoZero (algo), DeepSpeech2 (ds2), FasterRCNN (fast), NCF\_recommendation (ncf), Sentimental\_seqCNN (sent), Transformer\_fwd (trf), Yolo\_tiny (yolo). These models are selected from diverse machine learning domains, such as computer vision, speech recognition, natural language processing, gaming, and personalized recommendation.

\noindent\textbf{Memory Protection Simulation}. We implemented accelerators based on Intel SGX, MGX, and $\Design$ security mechanisms, using an unprotected accelerator as a benchmark, and set the size of protected memory to 16GB. SGX uses a multi-level Integrity Tree with 56-bit VNs and 64-bit MACs, along with a 16KB VN cache and 8KB MAC cache, both using an LRU replacement policy for write-back and write-allocate strategies. We use two different protection granularities: 64B and 512B. To ensure fairness, $\Design$ stores layer MACs off-chip.

\begin{table*}[t]
    \caption{Comparison of Memory Protection Schemes.}
    \label{table:comparison}
    \centering
    \footnotesize
    \renewcommand{\arraystretch}{0.85}
    \setlength{\tabcolsep}{1.8pt}
    \begin{tabular}{c|c|c|c|c|c}
        \toprule
        \textbf{Protection Scheme} & \textbf{Encryption Granularity} & \textbf{Integrity Granularity} 
                   & \textbf{Off-chip Memory Access} & \textbf{DNN Tiling Pattern} & \textbf{Encryption Scalability} \\ 
        \midrule
        \textbf{SGX-64B} & 16B & 64B & MAC,VN,IT  & \textcolor{pink}{\large \ding{55}}  &  \textcolor{pink}{\large \ding{55}} \\
        \rowcolor[HTML]{EFEFEF}
        \textbf{SGX-512B} & 16B & 512B & MAC,VN,IT & \textcolor{pink}{\large \ding{55}} &  \textcolor{pink}{\large \ding{55}} \\
        \textbf{MGX-64B} & 16B & 64B & MAC &\textcolor{pink}{\large \ding{55}} & \textcolor{pink}{\large \ding{55}} \\
        \rowcolor[HTML]{EFEFEF}
        \textbf{MGX-512B} & 16B & 512B & MAC & \textcolor{pink}{\large \ding{55}} & \textcolor{pink}{\large \ding{55}} \\
        \rowcolor[HTML]{DAE8FC}
        \textbf{\Design} & bandwidth-aware & multi-level & minimal to no cost & \textcolor{green}{\large \cmark} & \textcolor{green}{\large \cmark} \\
        \bottomrule
    \end{tabular}
    \begin{flushleft}
        \small
        \hspace{0.6 cm} *Note: "IT" signifies integrity tree.
    \end{flushleft}
\end{table*}

\begin{figure}
    \centering
    \includegraphics[width=0.92\linewidth]{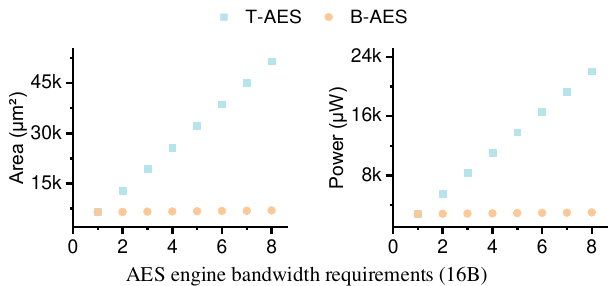}
    \caption{The area and power with increasing AES engine bandwidth requirements.}
    \label{figs:eval_area_power}
\end{figure}

\begin{figure}[t]
    \centering
    \includegraphics[width=0.8\linewidth]{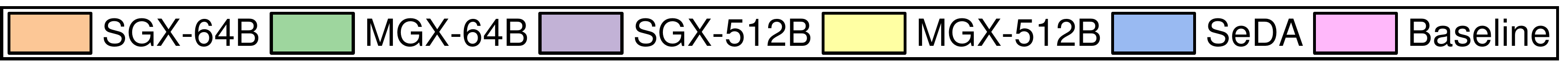}\\
    \vspace{-2mm}
    \subfigure[Server NPU]{\includegraphics[width= 0.98\linewidth]{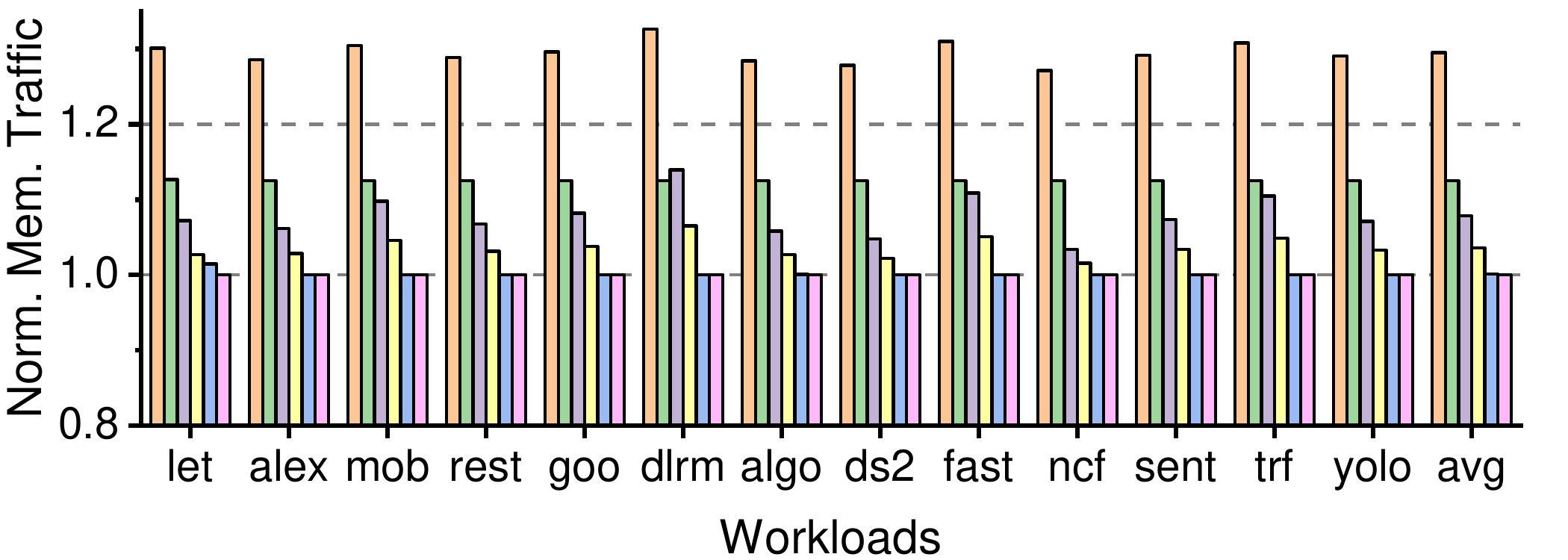}}\\
    \vspace{-3mm}
    \subfigure[Edge NPU]{\includegraphics[width= 0.98\linewidth]{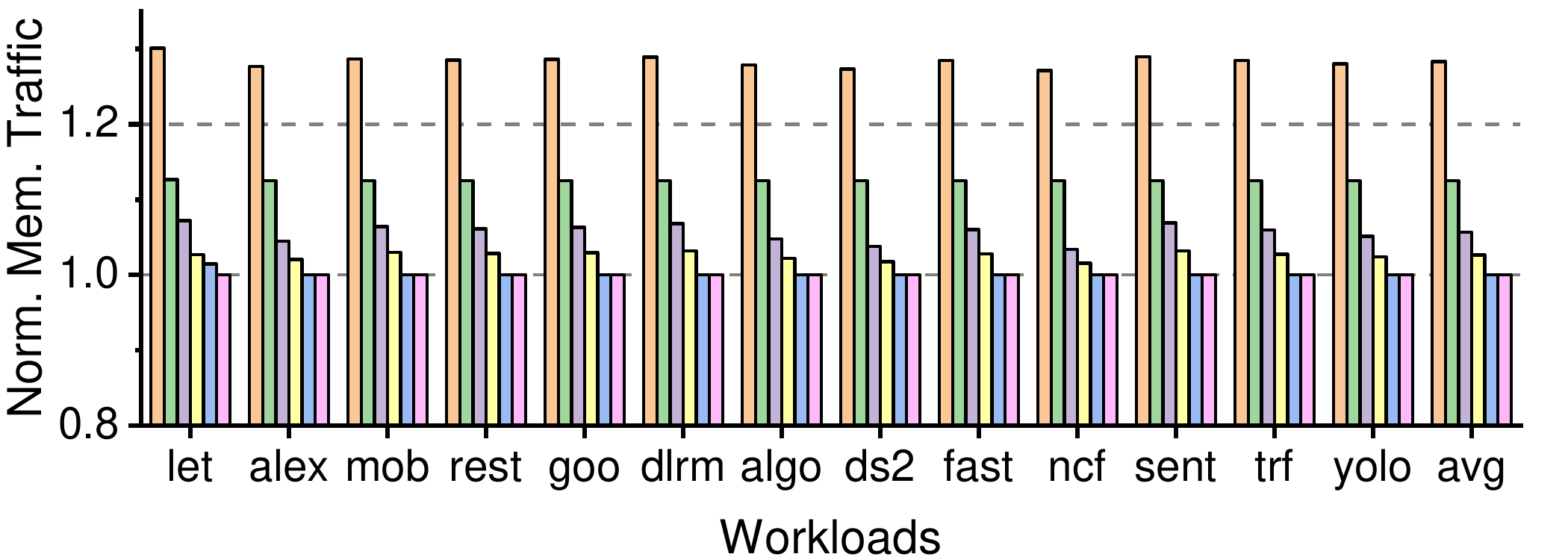}}
    \vspace{-2mm}
    \caption{The normalized memory traffic of memory protection schemes for various workloads.}
    \label{figs:eval_traffic}
\end{figure}

\subsection{Experimental Results}
We compare the memory traffic, which refers to the data exchanged between off-chip memory and accelerators, and performance across a unprotected baseline and five protection schemes: SGX-64B, SGX-512B, MGX-64B, MGX-512B, and our proposed $\Design$, as shown in Table~\ref{table:comparison}. All results are normalized to the baseline without protection.

\textbf{Area and Power.}
We developed a simulator based on 28nm technology to evaluate area and power, utilizing the AES engine implementations detailed in~\cite{banerjee2017energy}. We refer to the bandwidth-aware encryption mechanism as B-AES, and other traditional methods using multiple AES engines as T-AES. Through simulation, we assessed the changes in area and power. The x-axis represents the bandwidth required by the encryption engine to match the accelerator's bandwidth needs, expressed as multiples of the bandwidth provided by a single AES engine. As shown in Fig.~\ref{figs:eval_area_power}, our proposed B-AES demonstrates strong scalability, with minimal increases in area and power consumption as the accelerator bandwidth increases.

\textbf{Memory traffic.} 
Fig.~\ref{figs:eval_traffic} compares the memory traffic overhead using the DNN simulation configurations listed in Table~\ref{table:config}. On average, SGX-64B increases memory traffic by $30\%$ for Server NPU and $28.29\%$ for Edge NPU. In contrast, MGX-64B, without the overhead from additional VNs and MT, increases traffic by only $12.51\%$ and $12.63\%$, respectively. Increasing the protection granularity from 64B to 512B significantly reduces memory traffic. SGX-512B cuts traffic by $7.83\%$ on Server NPU and $5.13\%$ on Edge NPU. MGX-512B achieves reductions of $3.59\%$ and $2.39\%$ on these devices, respectively. Expanding the protection granularity to 512 bytes reduces memory traffic by minimizing security metadata. However, using larger data blocks can lead to inefficiencies due to misalignment with intra-layer tiling overlaps and varying inter-layer tiling patterns. This mismatch can hinder memory access and resource utilization, emphasizing the need for a balanced approach in selecting granularity to improve system performance and efficiency. Our proposed $\Design$ uses a multi-level integrity verification mechanism by XORing all optBlk MACs into a layer MAC. Fig.~\ref{figs:eval_traffic} shows that $\Design$ introduces near-zero memory traffic, with an overhead of only $0.12\%$ for Server NPU and $0.03\%$ for Edge NPU, highlighting the superiority of our proposed protection mechanism.

\begin{figure}[t]
    \centering
    \includegraphics[width=0.8\linewidth]{figures/eval_legend.pdf}\\
    \vspace{-2mm}
    \subfigure[Server NPU]{\includegraphics[width= 0.98\linewidth]{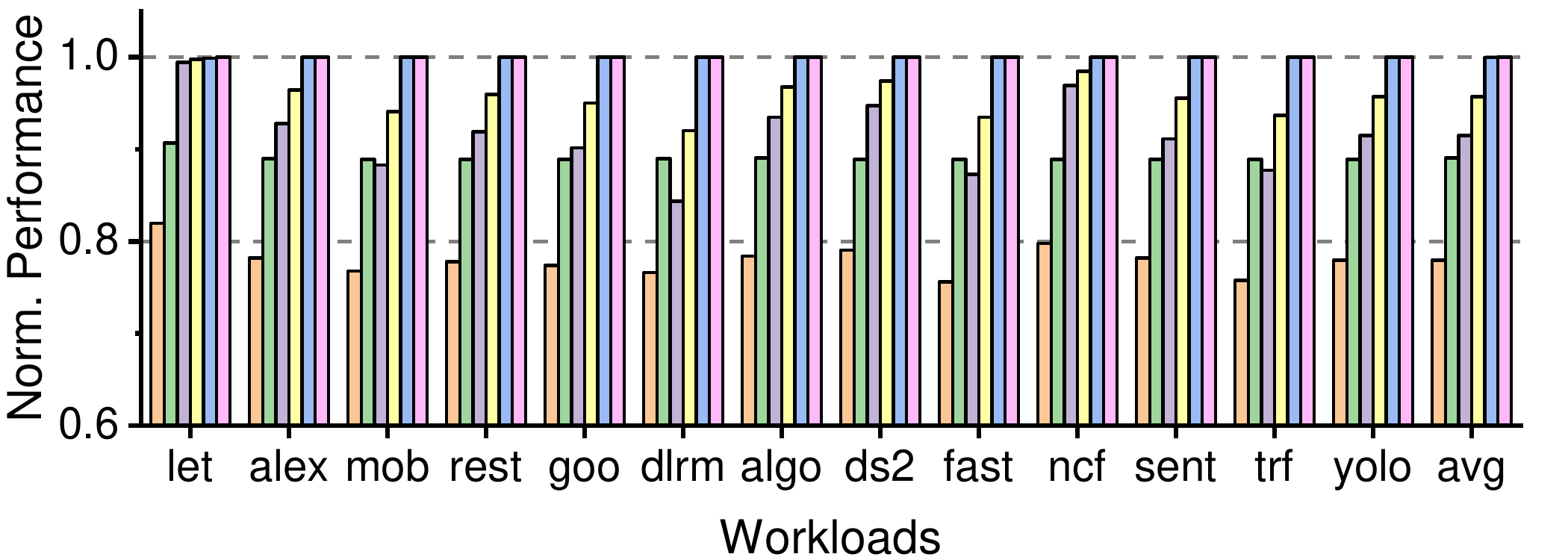}}\\
    \vspace{-3mm}
    \subfigure[Edge NPU]{\includegraphics[width= 0.98\linewidth]{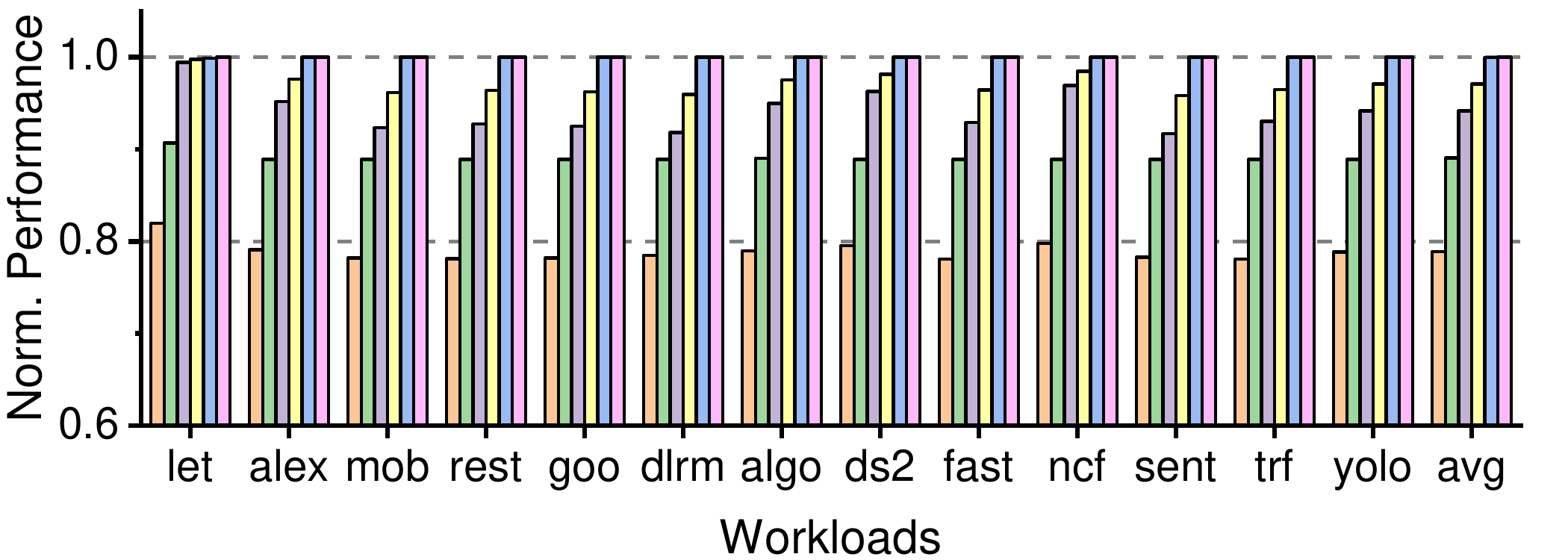}}
    \vspace{-2mm}
    \caption{The normalized performance of memory protection schemes for various workloads.}
    \label{fig:performance}
\end{figure}

\textbf{Performance.} Fig.~\ref{fig:performance}(a) presents a performance comparison analysis among the baseline and five memory protection mechanisms on Server NPU. SGX-64B is $22.04\%$ slower, MGX-64B is $10.93\%$ slower, SGX-512B is $8.49\%$ slower, and MGX-512B is $4.28\%$ slower than the unprotected baseline. In contrast, our proposed $\Design$ impacts performance by less than $1\%$, making it nearly negligible. This is primarily attributed to proposed multi-level integrity verification mechanism, which can reduce or even completely eliminate the performance overhead of off-chip memory accesses caused by integrity verification. By minimizing the storage and retrieval of MACs in off-chip memory, $\Design$ effectively manages the security metadata overhead from integrity verification.  Additionally, by storing these small layer or model MACs directly in on-chip SRAM, we can remove the performance overhead associated with off-chip access for integrity verification. For Edge NPU, Fig.~\ref{fig:performance}(b) presents similar findings. Compared to the baseline, SGX-64B, MGX-64B, SGX-512B, and MGX-512B slow down by $21.10\%$, $10.95\%$, $5.84\%$, and $2.90\%$, respectively. Remarkably, our proposed $\Design$ results in an almost imperceptible performance drop.
\section{Conclusion} \label{sec:conclusion}

In this work, we introduce $\Design$, a secure and efficient DNN accelerator architecture designed to ensure confidentiality and integrity in untrusted environments. Using bandwidth-aware encryption and multi-level integrity verification mechanisms, $\Design$ provides excellent scalability with minimal overhead to match computational bandwidth of accelerators, and significantly reduces or even eliminates off-chip memory access overhead. Experimental results show that proposed $\Design$ meets bandwidth requirements with near-zero hardware overhead, and significantly reduces the performance overhead compared to the state-of-the-art approaches.



\bibliographystyle{IEEEtran}
\bibliography{reference}
\end{document}